\documentclass[%
 reprint,
 amsmath,amssymb,
 aps,
]{revtex4-2}

\usepackage{graphicx}% Include figure files
\usepackage{dcolumn}% Align table columns on decimal point
\usepackage{bm}% bold math
\begin{document}

\preprint{APS/123-QED}

\title{Analysis of the PICUP Collection: Strengths and Areas for Development}% Force line breaks with \\

\author{W. Brian Lane}
%  \altaffiliation{Department of Physics, University of North Florida.}%Lines break automatically or can be forced with \\
 \email{Brian.Lane@unf.edu}
\affiliation{%
 Department of Physics, University of North Florida\\
 1 UNF Drive, Jacksonville, FL, 32224% \textbackslash\textbackslash
}%

% \collaboration{MUSO Collaboration}%\noaffiliation

%  \homepage{http://www.Second.institution.edu/~Charlie.Author}
% \affiliation{
%  Second institution and/or address\\
%  This line break forced% with \\
% }%
% \affiliation{
%  Third institution, the second for Charlie Author
% \affiliation{%
%  Authors' institution and/or address\\
%  This line break forced with \textbackslash\textbackslash
% }%

% \collaboration{CLEO Collaboration}%\noaffiliation

\date{\today}% It is always \today, today,

\begin{abstract}
The PICUP Collection of Exercise Sets (\href{https://www.compadre.org/PICUP/exercises/}{https://www.compadre.org/PICUP/exercises/}) contains over 60 peer-reviewed computation-infused activities for use in various physics courses from high school through graduate study. Each Exercise Set includes an instructor guide, student-facing exercises, and sample implementations for one or more programming platforms. We present an analysis of this Collection based on Exercise Set traits such as course level, word count, completion time, course context, computational methods, programming platforms, and engagement elements. This analysis highlights the strengths of the PICUP Collection (such as variety in subject and method coverage) and where there is ample room for development (such as expansion of high school- and advanced-level offerings and a greater representation of programming platforms).
% \begin{description}
% Secondary publications and information retrieval purposes.
% You may use the \texttt{description} environment to structure your abstract;
% use the optional argument of the \verb+\item+ command to give the category of each item. 
% \end{description}
\end{abstract}

\maketitle

%\tableofcontents

\section{\label{sec:intro}the picup collection}

Computation has become an invaluable facet of physics education, both to help students learn physics concepts \cite{AAPTRECS}, and to help students develop computational research skills in high demand \cite{GL20}. Computation offers many pedagogical benefits, such as bridging between analytical and experimental approaches \cite{weller2018investigating}, reducing the barrier of mathematical accessibility \cite{symbolic}, opening analytically intractable problems to study \cite{rev}, and engaging students with novel means of visualization \cite{qPhET}. Harnessing these benefits requires instructors to (1) consider students' prior programming experiences (or the lack thereof) and their comfort level or apprehension related to programming \cite{bioExcel}; (2) evaluate programming environments, tools, and techniques for accessibility and engagement \cite{fernandez2017improving}; (3) minimize cognitive overhead for novice programmers \cite{whalley2014qualitative}; and (4) distill unique meaningful learning from each computational activity while connecting computation as a practice to theory and experiment \cite{weller2018investigating}. No singular physics educator has the time or expertise to fully engage in all of these tasks throughout each course they teach, and yet it is increasingly important that we continue to integrate computation more deeply into our instruction \cite{bott2020student}. Physics educators can ease this burden by sharing computational resources.

For this reason, the Partnership for Integration of Computation into Undergraduate Physics (PICUP) was formed ``to increase the use of computation in physics courses through the development of computational packages that are closely tied to the subject matter of popular physics texts, and through lowering the barriers to the adoption, and more importantly, adaptation, of educational materials so that the integration of computation into physics courses is effectively facilitated for all physics faculty'' \cite{PICUP:main}. PICUP accomplishes this goal by hosting professional development workshops and webinars, organizing conference sessions, maintaining a professional network of physics educators committed to the incorporation of computation into the physics curriculum, and hosting a collection of peer-reviewed exercise sets \cite{PICUP-COP}.

The PICUP Collection of Exercise Sets \cite{collection} features computation-infused learning activities freely available for educators to adopt and adapt in their courses. These Exercise Sets (ESs) are authored by physics educators and then peer reviewed by PICUP editors and referees who have previously published at least one ES. Each ES is required to follow a template (see below) and is peer-reviewed based on educational quality, consistency in content and format, appropriateness for the intended curriculum, and correctness of the physics and computational content \cite{reviewprocess}. (PICUP also hosts a Faculty Commons page of educator-developed resources, but these are not subject to peer review and are not considered here \cite{FacultyCommons}.) At the time of this article's submission, the PICUP Collection includes 62 ESs. This manuscript's author has published two such ESs and peer reviewed one published ES.

The required template for an ES includes each of the following sections. When an ES is published, the Exercises section is visible to anyone visiting the PICUP web site, but all other sections require a verified instructor login through ComPADRE.
\begin{itemize} 
\item \textit{Description} This (usually brief) summary helps instructors evaluate whether an ES is of interest for their courses.
\item \textit{Learning Objectives} This list identifies what the students will be able to do by the end of the ES. Each learning objective is expected to include the Exercise(s) in which the learning objective is accomplished.
\item \textit{Instructor's Guide} This summary provides the instructor an overview of the ES and includes recommendations for in-class deployment.
\item \textit{Theory} This section outlines the foundational physics concepts employed in the ES.
\item \textit{Experiments} This optional section includes information about any experiments that accompany the ES.
\item \textit{Exercises} These student-facing instructions and questions can be distributed directly to students or adapted by the instructor as needed.
\item \textit{Pseudocode} Here, an ES author can present an outline of the code to be developed or used in the ES. Using pseudocode to engage students in the process of planning a code has shown to be effective in promoting computational thinking \cite{davies2008effects}.
\item \textit{Code Templates} At the instructor's discretion, these starter codes or minimally working programs \cite{weatherford2013mwp,oleynik2019mwp} can be provided to students as a place to begin the ES, instead of requiring students to develop their programs from scratch, which can be costly in terms of time and student motivation. A code template is an example of ``webbing,'' supporting student learning by providing a structure that students can extend and reconstruct as needed \cite{langbeheim2020webbing,noss1996webbing}.
\item \textit{Solutions} These sample answers to the Exercises help the instructor form expectations for the ES and grade student submissions.
\item \textit{Completed Code} This example represents the final product the students should develop during the ES.
\end{itemize}

\begin{figure}[b]
\includegraphics[width=0.45\textwidth]{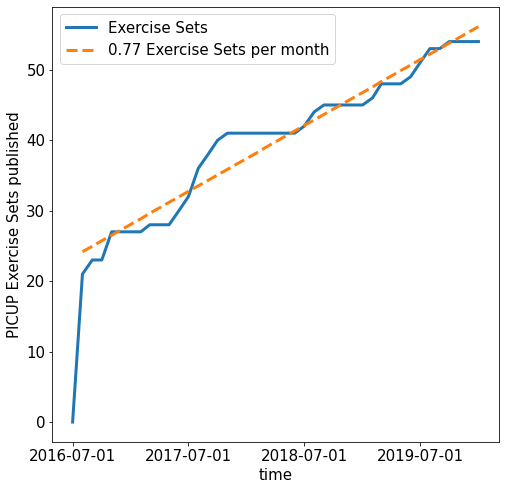}
\caption{\label{fig:ESvsTime} Growth of the PICUP Collection of Exercise Sets. After the Collection's initial launch with 20 Exercise Sets, PICUP has published an average of 0.77 Exercise Sets each month.}
\end{figure}

The PICUP Collection launched in summer 2016, with a debut publication of 20 ESs. Since then, PICUP has published an average of 0.77 new ESs each month, as shown in Figure \ref{fig:ESvsTime}. These ESs are written by 31 authors, 13 of whom have published more than one ES. Nearly all (58 out of 62) ESs are written by a single author, and none is written by more than 2 authors. An internet search identifies all ES authors as being affiliated with institutes of higher education or national AAPT leadership.

As the demand for computational assignments increases and the PICUP Collection continues to grow in size, it is worthwhile to evaluate the strengths and areas for development in the PICUP Collection. This article carries out such an evaluation by examining the properties of the current body of PICUP ESs.

\section{Evaluation Framework}

To carry out this evaluation, we adopt the following framework. We assume that an instructor considers an ES for adoption in a particular course context, described by a course level (representing the general preparedness of the students), a general subject (the course being taught), and one or more topics (unit-level content that addresses a subset of the course's learning objectives) that they are seeking an ES to address. For the sake of efficiency, instructors are likely to select an ES that most specifically caters to these characteristics in their course context rather than select an ES that requires significant adaptation for their context. For example, a high school physics teacher looking for an ES about standing waves and a university professor looking for an ES about time evolution in quantum mechanics are evaluating ESs with very different criteria, even if the ESs about these subjects might involve similar principles (such as waves and boundary conditions).

We consider the conceptual content of an ES to be complemented by one or more computational methods that enable students to investigate a topic as directed in the ES. By ``computational method,'' we mean a procedure or algorithm implemented in a code to generate a numerical solution or simulate a physical problem. A computational method is usually based on a mathematical derivation or principle and might be used across multiple subject contexts. In the design of an ES, a computational method is usually chosen because it offers a numerical solution to a problem that is out of reach mathematically or experimentally, such that the ES enables students to explore a wider range of physical scenarios. Depending on an ES's learning objectives, the computational method might receive minimal attention within the ES (for example, to focus students' time on the physics concepts being illustrated), or the computational method might be the ES's primary focus (for example, to explicitly train students in implementing and extending a broadly useful method). In principle, an ES can employ any number of computational methods in combination.

We also assume that instructors consider which programming platform they will use when delivering an ES. By ``programming platform,'' we mean the programming language used and the computer-interactive environment in which the student uses code written in that language. We recognize that an instructor's choice of programming platform is determined by many pragmatic factors, such as the instructor's familiarity, student preferences, and institutional availability or support, and that these factors might easily conflict with each other \cite{bioExcel}. However, we also agree with the common sentiments within the PICUP community that (1) most computation-based learning goals transcend one's choice of platform, (2) ideally, an ES should be implementable in any platform, and (3) apart from specific programmatic needs, any platform can be an appropriate choice.

Finally, we assume that many ESs are designed with the goal of engaging students in analysis and reflection beyond simply running code. An ES might accomplish this goal by prompting students to analyze results, reflect on their findings, or synthesize multiple sources of learning. Such engagement might include insightful forms of data visualization or integration between computation and experiment.

Given these assumptions, we suggest that \textit{variety} should be a goal for the PICUP Collection, both in the features outlined above (course contexts, computational methods, programming platforms, and higher-level engagement) and in the frequency with which those features are found in combination across the Collection. With this goal in mind, this paper proceeds by reviewing the features and contents of the ESs within the PICUP Collection and examining the frequency with which common features occur. In doing so, we will answer questions such as: What difficulty level do the ESs typically pose? What subjects are well-served by the Collection? What computational methods are most frequently implemented? What programming platforms are most frequently employed? In what ways are ESs designed to engage students in higher-level thinking? What combinations of these characteristics occur most frequently? What sorts of ESs are needed to strategically expand the Collection? 

\section{Challenge Level}

One of the most important qualities of an instructional assignment is its appropriateness for the intended audience. This factor is especially important to consider in a computational activity, given that programming is known to bring about apprehension in students \cite{bioExcel} (especially first-time programmers), which compounds the apprehension that many students already feel toward physics \cite{apprehension}. In this section, we examine the PICUP Collection's ESs based on their intended audience (Course Level), estimated completion time, and word count.

\subsection{Course Levels}

PICUP ESs are required to list at least one Course Level, representing a general location in the physics curriculum where the ES can appropriately be used. The PICUP authoring interface presents four options for Course Level: High School, First Year, Beyond the First Year, and Advanced. While the first two Course Levels are clearly distinguished, the last two (Beyond the First Year and Advanced) are perhaps less clearly delineated (Does Advanced mean senior undergraduate, graduate, or both?) and their distinction is negotiated during the review of each ES. An author can select multiple Course Levels as appropriate; for example, nearly any ES designed for High School is likely appropriate for a First Year context, though not necessarily vice versa.

\begin{figure}[tb]
\includegraphics[width=\columnwidth]{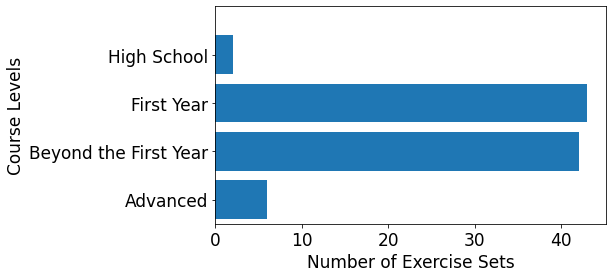}
\caption{\label{fig:CourseLevelsBarGraph} Distribution of PICUP Exercise Sets by Course Level. The majority of Exercise Sets are intended for First Year and Beyond the First Year contexts.}
\end{figure}

Figure \ref{fig:CourseLevelsBarGraph} shows the number of Exercise Sets tagged with each Course Level. (The total number exceeds the $N = 62$ ESs since many ESs are tagged with multiple Course Levels.) The vast majority of ESs are designed for a First Year or Beyond the First Year context, while currently only a handful are designed for High School (2 ESs) or Advanced (6 ESs) contexts. Such a concentration in the core of the undergraduate experience is to be expected, given PICUP's explicit focus on enhancing undergraduate education and the lack of high school teachers among ES authors. Indeed, the High School designation was only recently added to the ES submission options, so it may be possible that some First-Year ESs published before the High School designation was available are appropriate for a High School context, as well.

\begin{figure}[tb]
\includegraphics[width=0.45\textwidth]{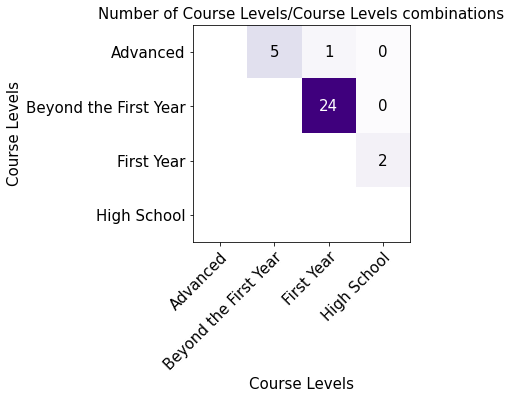}
\caption{\label{fig:Level-Level} Overlap in PICUP Exercise Set Course Levels. The Values indicate the number of Exercise Sets that list both Course Levels. Over one-third of the ESs list First Year and Beyond the First Year as Course Levels.}
\end{figure}

An ES can list multiple Course Levels, and we see in Figure \ref{fig:Level-Level} that 24 out of the 62 total ESs list both the First Year and Beyond the First Year levels. Nearly all of the ESs listed as Advanced are also listed as Beyond the First Year, perhaps confirming the ambiguity between the two levels noted above. Both High School ESs are also listed as First Year-appropriate. Only one ES (``Computing the 1-D Motion of a V2 Rocket'') lists 3 Course Levels (First Year, Beyond the First Year, and Advanced), and no ESs list all 4 Course Levels.

\subsection{Completion Time}

\begin{figure}[tb]
\includegraphics[width=0.45\textwidth]{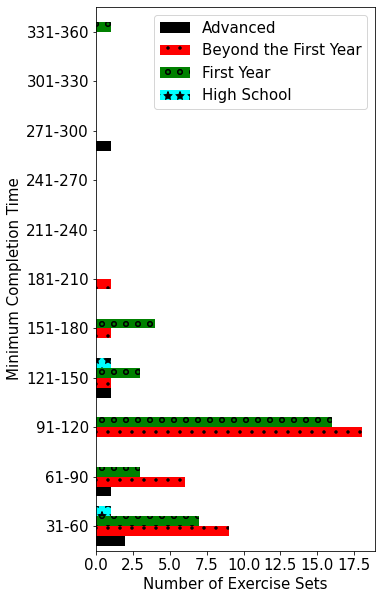}
\caption{\label{fig:CompletionTimeBarGraph} Minimum completion time for PICUP Exercise Sets, organized by Course Level.}
\end{figure}

Another measure of the complexity of an assignment is its expected completion time. ES authors are asked to estimate the completion time required to help instructors properly allocate time and credit for the ES in their courses. Figure \ref{fig:CompletionTimeBarGraph} shows these estimated times for ESs as grouped by Course Level. (Some ESs list a range of completion times, in which case only the minimum value is represented here.) Most ESs can be completed in 2 hours or less. This trend is likely due to authors allocating extended lab time to the completion of computational activities, as is common practice.

\subsection{Word Count}

\begin{figure}[tb]
\includegraphics[width=0.45\textwidth]{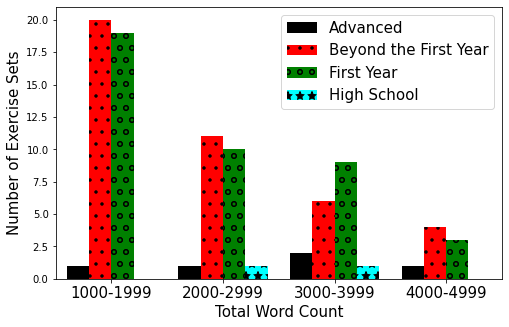}
\caption{\label{fig:WordCountBarGraph} Total word count for for PICUP Exercise Sets, organized by Course Level.}
\end{figure}

Finally, we discuss the word count of each ES, totaled across the various sections discussed in Section \ref{sec:intro} (except for code templates and completed code). The word counts for these ESs are shown in Figure \ref{fig:WordCountBarGraph} grouped by Course Level.

Nearly one-third of the First Year and Beyond the First Year ESs fall below 1,000 words in total, and only a handful of ESs exceed 4,000 words. Interestingly, the two High School ESs fall in the middle range of word count, between 2,000 and 4,000, making them longer than many undergraduate-level ESs. This difference might be due to the additional scaffolding required to implement computational physics in a high school context.

\section{content features}

In this section, we examine the conceptual and technical contents to be found in PICUP ESs, grouped by Course Level. By ``contents,'' we mean course-level subjects, unit-level topics, computational platforms, the balance between traditional and computation-based learning objectives, and elements that might enhance student engagement.

\subsection{Course Contexts and Topics}

When submitting an ES to the PICUP Collection, authors are asked to select one or more Course Contexts from among Astronomy/Astrophysics, Electricity \& Magnetism, Experimental Labs, Mathematical/Numerical Methods, Mechanics, Modern Physics, Other, Quantum Mechanics, Thermal \& Statistical Physics, Waves \& Optics, Condensed Matter Physics, High Energy/Particle Physics, Biophysics, and Chemical/Molecular Physics. (At the time of this article's writing, there were no ESs published for Condensed Matter Physics, High Energy/Particle Physics, Biophysics, or Chemical/Molecular Physics, so they are not discussed in this analysis.) These Course Contexts more or less map onto the courses an ES is likely to be adopted in, but academic programs might differ in where they place some topics in their curricula. For example, there is notable overlap in topics between Modern Physics and Quantum Mechanics, and two academic programs might organize topics differently under these two course headings. As with selecting Course Level, the distinction between these Course Contexts is negotiated during the ES review process. 

\begin{figure}[tb]
\includegraphics[width=0.5\textwidth]{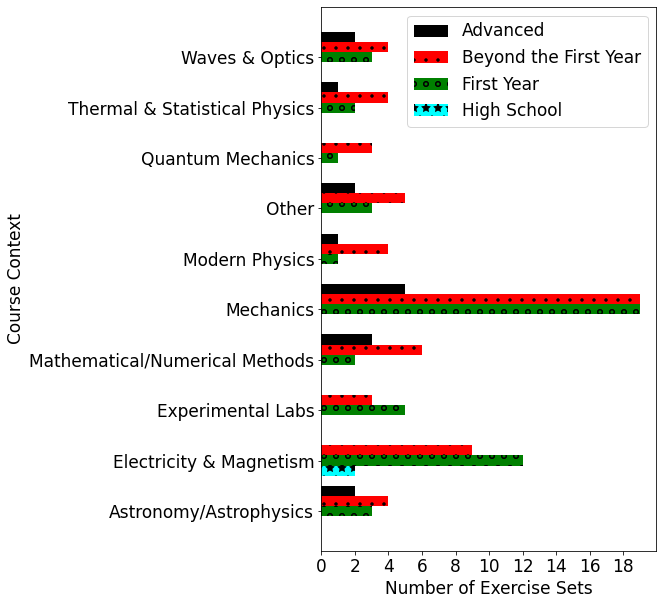}
\caption{\label{fig:SubjectAreaBarGraph} Distribution of Course Contexts across the PICUP Collection, organized by Course Level.}
\end{figure}

Figure \ref{fig:SubjectAreaBarGraph} shows the number of ESs tagged with each Course Context, organized by Course Level. Mechanics holds a definitive lead in the number of ESs, followed by Electricity \& Magnetism. This prominence makes sense, given the high number of First-Year ESs counted in Figure \ref{fig:CourseLevelsBarGraph}. However, this subject emphasis is found among Beyond the First Year ESs, as well, perhaps indicating that many ES authors design their ESs for the introductory context and cross-list them as Beyond the First Year with sophomore-level courses in mind. 

There is a notable lack of ESs among many subjects traditionally taught at the junior and senior level. Very few ESs address Modern Physics, Quantum Mechanics, and Thermal \& Statistical Physics. No ESs at the time this article was written address Condensed Matter Physics, High Energy/Particle Physics, Biophysics, or Chemical/Molecular Physics. There are no Electricity \& Magnetism ESs marked for an Advanced Course Level.

\begin{figure}[tb]
\includegraphics[width=0.5\textwidth]{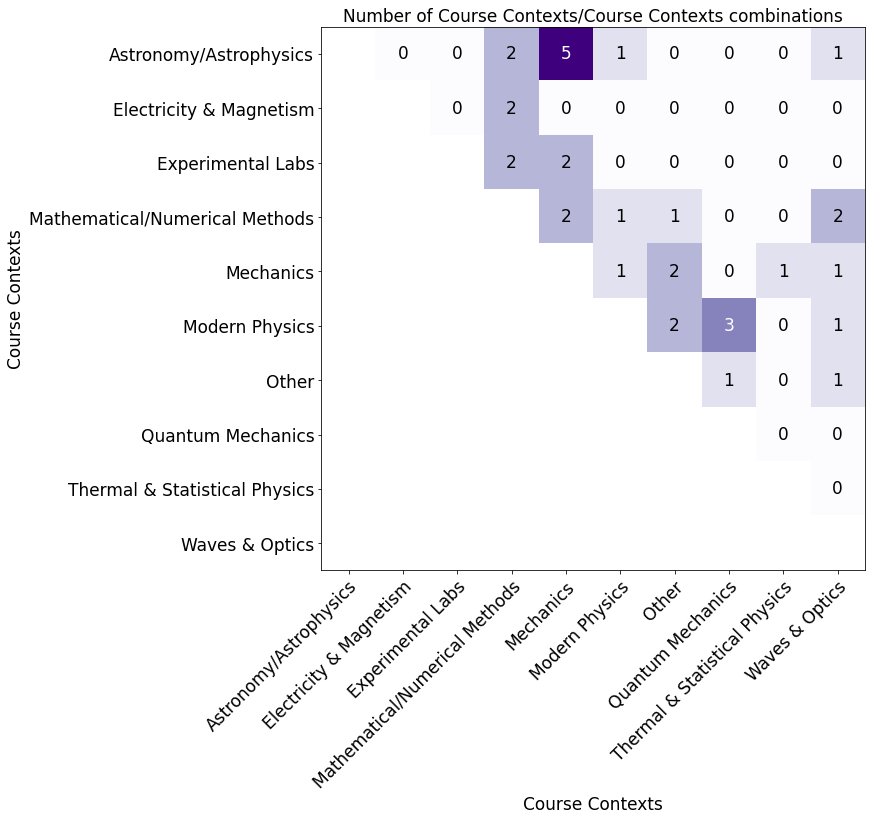}
\caption{\label{fig:SubjectHeatMap} Combinations of Course Contexts within PICUP Exercise Sets. Values indicate the number of Exercise Sets that list both Course Contexts.}
\end{figure}

Authors may select multiple Course Contexts as appropriate. For example, suppose an author developed an ES that studies the motion of two charged particles under the influence of their mutual electrostatic force. The author could tag this ES with Mechanics (since it involves the influence of force on motion), Electricity \& Magnetism (since it involves charged particles experiencing the Coulomb force), Thermal \& Statistical Mechanics (if the author extended the code to study many particles), and Condensed Matter Physics (if the author extended the code to study classical models of conduction). Figure \ref{fig:SubjectHeatMap} shows the frequency of Course Context cross-selection among ESs by counting the number of ESs tagged with each pairwise combination of Course Contexts. The most frequent combination (5 ESs) occurs between Astronomy/Astrophysics and Mechanics, which is reflective of the high number of gravitation-based ESs. The second most frequent combination (3 ESs) occurs between Modern Physics and Quantum Mechanics, reflecting the topics frequently taught in common between the two subjects. The remaining combinations of Course Contexts occur infrequently (0-2 times).

The Course Contexts listed in the PICUP Collection organize ESs by the general course contexts in which they might be used, but it is also useful to examine the specific topics addressed by each ES. By ``topics,'' we mean unit-level concepts that students might encounter in an individual chapter or section of a course textbook. A topic might be a broadly useful principle such as conservation of energy or a specific application such as drag force. We have identified topics with as fine a grain as possible, listing multiple topics in an ES when applicable. For example, 4 ESs that specifically study the motion of a pendulum were identified as having topics ``pendulum'' and ``rotational motion,'' since pendulum motion is a specific example of rotational motion.

\begin{figure}[tb]
\includegraphics[width=0.45\textwidth]{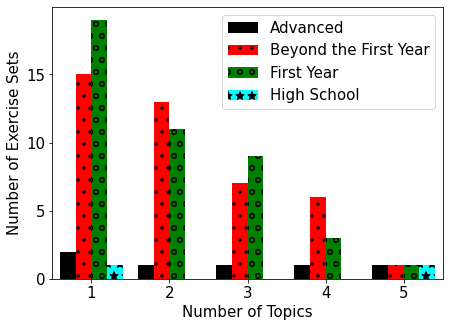}
\caption{\label{fig:TopicsCountBarGraph} Number of topics found in each PICUP Exercise Set, organized by Course Level.}
\end{figure}

Our analysis identified 60 unique topics among the 62 ESs. For the sake of space, we will not list the full set of topics here, but instead discuss the number of topics addressed in each ES and comment on frequently addressed topics. As shown in Figure \ref{fig:TopicsCountBarGraph}, most ESs address 2 to 4 topics and none address more than 5 topics. The topics that occur most frequently are universal gravitation (11 ESs), electric fields of stationary charges (9 ESs), rotational motion (6 ESs), energy conservation (6 ESs), electric potential (5 ESs), and drag force (5 ESs). The remaining topics occur in 4 or fewer ESs. We observe from this trend that there is a great variety of physics topics represented in the PICUP Collection, with a few clear favorites that might occur in an introductory context.

\subsection{Computational Methods}

Computation is often incorporated into the physics curriculum with two primary goals in mind: To use computational activities to enhance the learning of physics concepts, and to teach students the use of established computational methods. By ``computational method,'' we mean a procedure or algorithm implemented in a code to generate a numerical solution or simulation of a physical problem. A computational method is usually based on a mathematical derivation or principle and might be used across multiple subject contexts. For example, numerical integration is a computational method based on the Riemann sum and is used in multiple subject contexts, such as finding center of mass, evaluating the electrostatic field produced by a charge distribution, and normalizing a wave function.

We identified 20 distinct computational methods used in these ESs. Some methods involve the use of external libraries as a ``black box'' (such as a differential equation solver in MATLAB or special functions from SciPy), while others implemented extended programming structures (such as a for or while loop in bracket root-finding or the Euler-Cromer method). We attempted to group together methods that differed only in technical details while maintaining distinctions between qualitatively different processes. For example, we grouped together any ESs that used the original Euler method or the modified Euler-Cromer method, as the only operational difference is the ordering of the update statements. On the other hand, we kept Euler-Cromer distinct from numerical integration, even though Euler-Cromer is a form of numerical integration, as the setup and type of problems that each method can address are qualitatively different.

Some PICUP ESs do not implement a computational method. For example, ``Lab Skills: Converting file formats'' is designed to train students in converting data files between formats frequently used in a lab setting, and ``Rainbows'' and ``Gravitational Waves from Binary Orbits'' focus on helping students work with the visualization of a physics principle. Data visualization is an important emerging skill \cite{visualization}, and we examine its presence across the PICUP Collection more fully in Subsection \ref{subsec:ee}.

\begin{figure}[tb]
\includegraphics[width=0.45\textwidth]{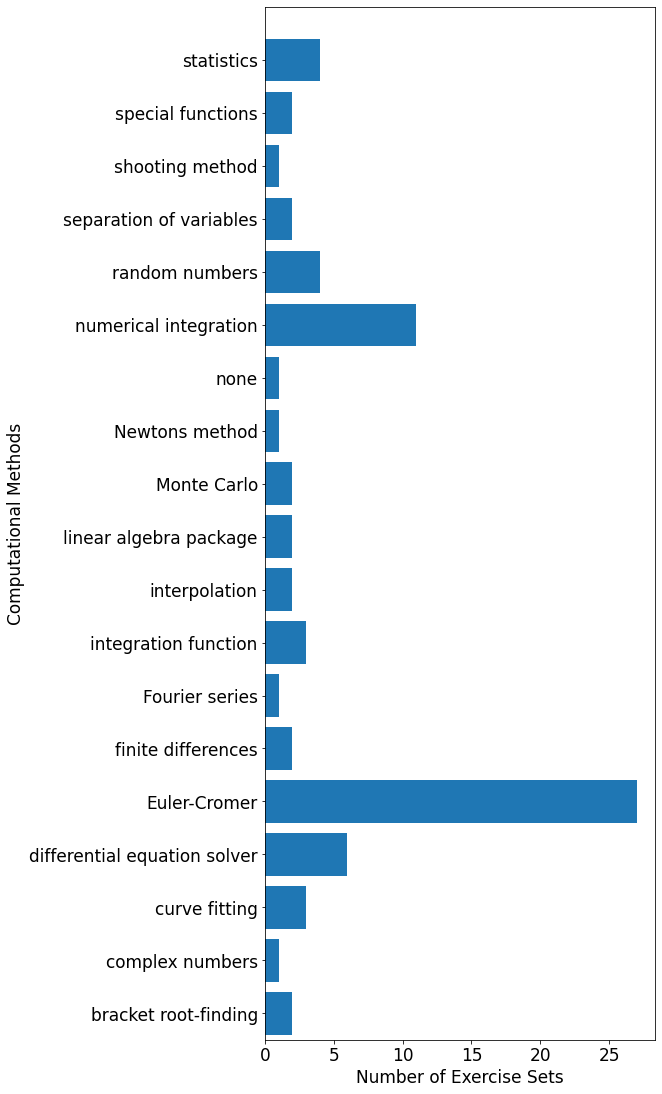}
\caption{\label{fig:CompMethbargraph} Distribution of computational methods employed in PICUP Exercise Sets.}
\end{figure}

Figure \ref{fig:CompMethbargraph} shows the number of ESs that use each of the 20 computational methods that we identified. The two most frequently used computational methods are the Euler-Cromer method and numerical integration. The Euler-Cromer method is frequently used in Mechanics contexts as a simple method of numerically solving Newton's Second Law for position as a function of time, and numerical integration is frequently used in Electricity \& Magnetism contexts to evaluate integrals over charge distributions. Unsurprisingly, the prominence of these two methods correlates with the high number of ESs focused on these two Course Contexts at the First Year level.

\begin{figure}[tb]
\includegraphics[width=0.45\textwidth]{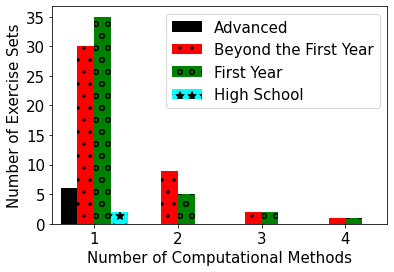}
\caption{\label{fig:MethCountbargraph} Number of computational methods employed in each PICUP Exercise Set.}
\end{figure}

In contrast to the frequent combining of multiple topics in an ES (Figure \ref{fig:TopicsCountBarGraph}), relatively few ESs implement multiple computational methods, as shown in Figure \ref{fig:MethCountbargraph}. This reliance upon a single computational method is likely due to the fact that these ESs may be a student's first introduction to these computational methods (if not their first experience with programming). With many students already apprehensive about completing a computational activity, limiting the number of computational methods required in an ES is important to the scaffolding process. It is interesting to note, however, that all of the Advanced-level ESs implement only a single computational method. One might expect the Advanced level to permit the integration of multiple computational methods. 

\subsection{Learning Objectives}

When selecting or developing a computational activity, instructors need to establish priorities regarding the development of students' conceptual understanding of physics and training students in the use of computational methods and practices. While these goals are certainly complementary and there is no ``wrong'' means of balancing them, on a pragmatic level, they compete for priority in time and assignment weighting. Each PICUP ES is required to include a list of learning objectives to help instructors understand the focus of each ES. 

We categorized each learning objective found in the Collection based on whether the objective addresses only physics learning, only computational training, or both. A learning objective that addresses physics learning might reference a physics concept or require students to interpret, validate, compare, or apply the results developed during the ES. A learning objective that addresses computational training might reference computational methods or software packages, visualization, or computing. We find that, of the 299 learning objectives presented across the Collection, 146 (48.8\%) address only physics learning, 135 (45.2\%) address only computational training, and 18 (6.0\%) address both. (We do not break down this evaluation by individual ES since the authors represented by the Collection do not necessarily take comparable approaches to writing learning objectives.) It seems, therefore, that the Collection as a whole presents a well-rounded balance between physics and computation. 

\subsection{Platforms}\label{subsec:platforms}

The selection of a programming platform (the programming language used and the computer-interactive environment in which the student uses code written in that language) is an important aspect of developing or adapting a computational activity. An instructor must balance their own familiarity, students' preferences and experience, and institutional resources \cite{bioExcel}. (Anecdotally, this process often proves non-trivial, and an investigation of best practices in platform selection would be beneficial.) When submitting an ES to the PICUP Collection, an author must include at least one code template and at least one completed code that fulfills the ES's requirements. Here, we examine the platforms that authors use in these codes.

\begin{figure}[tb]
\includegraphics[width=0.45\textwidth]{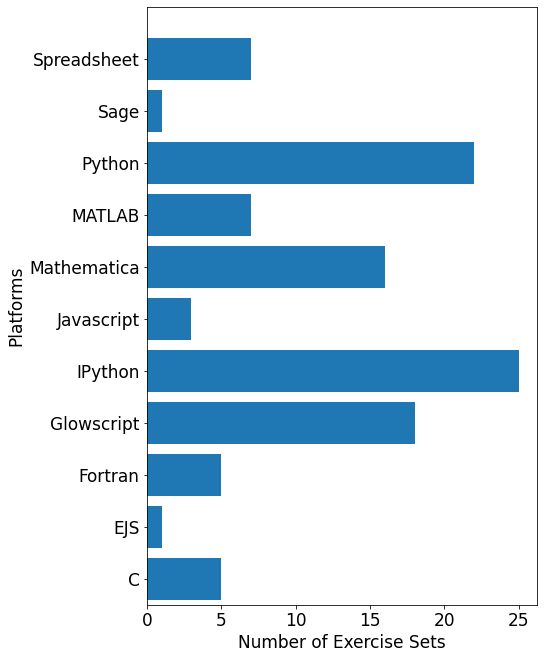}
\caption{\label{fig:Platformbargraph} Distribution of platforms employed in each PICUP Exercise Set.}
\end{figure}

We identified 11 different programming platforms among the 62 ESs in the PICUP Collection. Figure \ref{fig:Platformbargraph} shows the distribution of these platforms among the Collection. Note that some of these platforms may use the same programming language. For example, although the GlowScript, IPython, and Python platforms all use the same language, they are listed separately, since each offers different pedagogical opportunities. Following trends in physics and mathematics research, Python-based platforms are the most popular among the Collection, followed by Mathematica.

\begin{figure}[tb]
\includegraphics[width=0.45\textwidth]{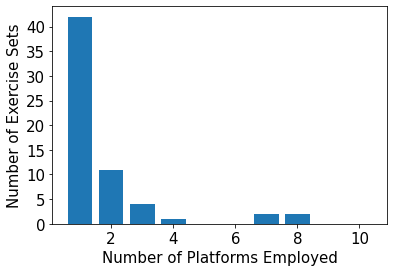}
\caption{\label{fig:PlatformCountbargraph} Number of platforms employed in each PICUP Exercise Set.}
\end{figure}

PICUP editors recommend that ESs use at least three different platforms \cite{reviewprocess}, as a greater number of readily available implementations makes an ES useful to more adopters. Figure \ref{fig:PlatformCountbargraph} shows the frequency with which ESs include code templates and completed codes with multiple platforms. We find that over two-thirds of the 62 ESs use only a single platform in their code templates and completed codes. Another handful use two platforms, and relatively few ESs use the recommended three or more.

\begin{figure}[tb]
\includegraphics[width=0.45\textwidth]{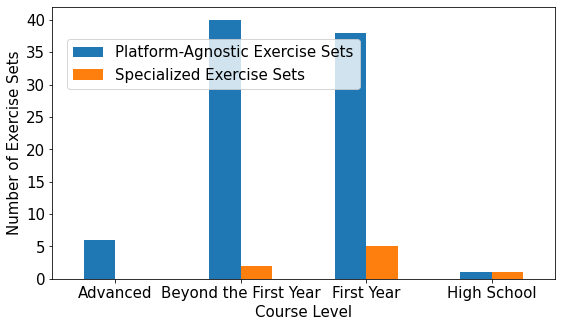}
\caption{\label{fig:AgnosticCountbargraph} Distribution of platform-agnostic and specialized Exercise Sets, grouped by Course Level.}
\end{figure}

Exercise Sets are also classified according to whether they are specialized for use with \textit{only} a single programming platform, or whether they are platform agnostic. This classification is different from the above-mentioned ESs that offer code templates and completed codes for only a single platform but \textit{could} be adapted to other platforms. For example, the ES ``Making Animations with Potential Energy'' relies on the three-dimensional animation elements of GlowScript so heavily that it would require a major overhaul to adapt this ES for other platforms. Figure \ref{fig:AgnosticCountbargraph} shows that, even though the majority of ESs include implementations in only a single platform, nearly all of the ESs (57 out of 62) are platform agnostic and could be extended to more platforms.

\subsection{Engagement Elements}\label{subsec:ee}

Finally, we examine the use of engagement elements across the PICUP ESs. By ``engagement elements,'' we mean features of the ES that take students beyond simply running code, and prompt them to analyze results, reflect on their findings, or synthesize multiple sources of learning. This definition includes basic visualizations such as line graphs and histograms and more advanced features such as three-dimensional animations and integration of computation with experiment.

\begin{figure}[tb]
\includegraphics[width=0.45\textwidth]{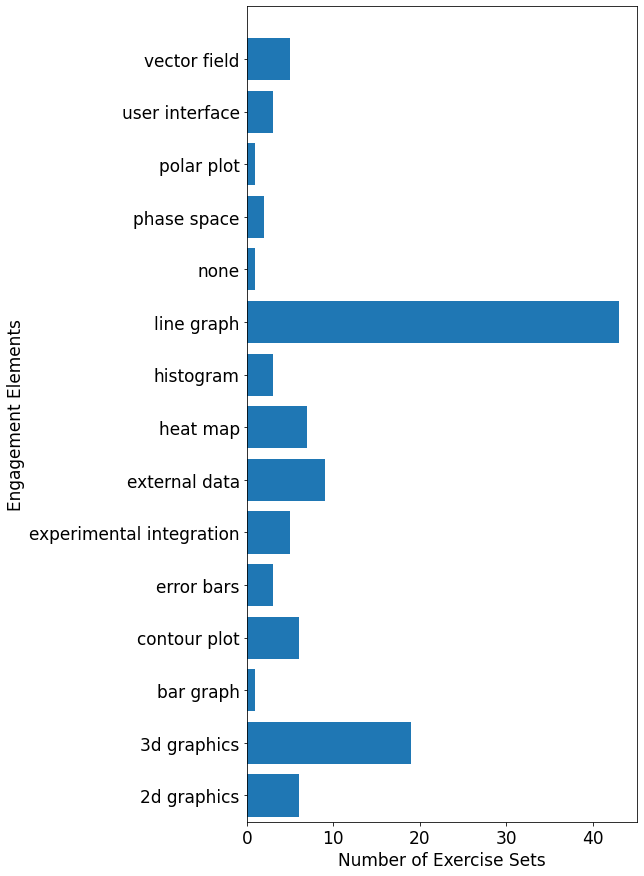}
\caption{\label{fig:EngagementElementbargraph} Distribution of engagement elements across Exercise Sets.}
\end{figure}

We identified 16 unique engagement elements in the PICUP Collection. Figure \ref{fig:EngagementElementbargraph} shows the number of ESs that use each of these engagement elements. The most common engagement element is a line graph, occurring in roughly two-thirds of the 62 ESs. This prevalence makes sense as most physics results are traditionally visualized as a line graph depicting the relationship between two quantities. The second most common engagement element is three-dimensional graphics, occurring in almost one-third of the ESs. Producing three-dimensional graphics has become much more accessible in recent years, and offers a notable learning advantage to the two-dimensional graphics found in physics textbooks.

\begin{figure}[tb]
\includegraphics[width=0.45\textwidth]{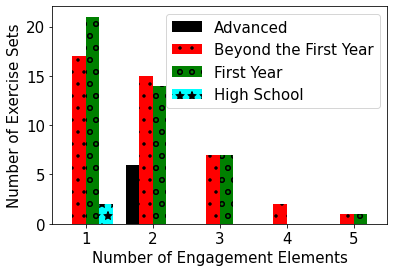}
\caption{\label{fig:EngagementElementCountbargraph} Number of engagement elements used in each Exercise Set, grouped by Course Level.}
\end{figure}

Figure \ref{fig:EngagementElementCountbargraph} shows the number of ESs that use multiple engagement elements. While it is most common for an ES to use a single engagement element, many ESs employ two or more. The most common combination (found in 9 ESs) is three-dimensional graphics and line graphs, followed by the combination of external data and line graphs (found in 7 ESs).

\section{Feature Combinations}

We next examine how the features discussed in the previous section (Course Contexts, computational methods, platforms, and engagement elements) are used in combination with each other throughout the PICUP Collection. Frequent combinations of content features indicate  a strong connection (such as between mechanics and the Euler-Cromer method or between electricity \& magnetism and numerical integration) while infrequent combinations indicate room for further development.

\subsection{Course Contexts and Computational Methods}\label{subsec:sacc}

\begin{figure*}[tb]
\includegraphics[width=\textwidth]{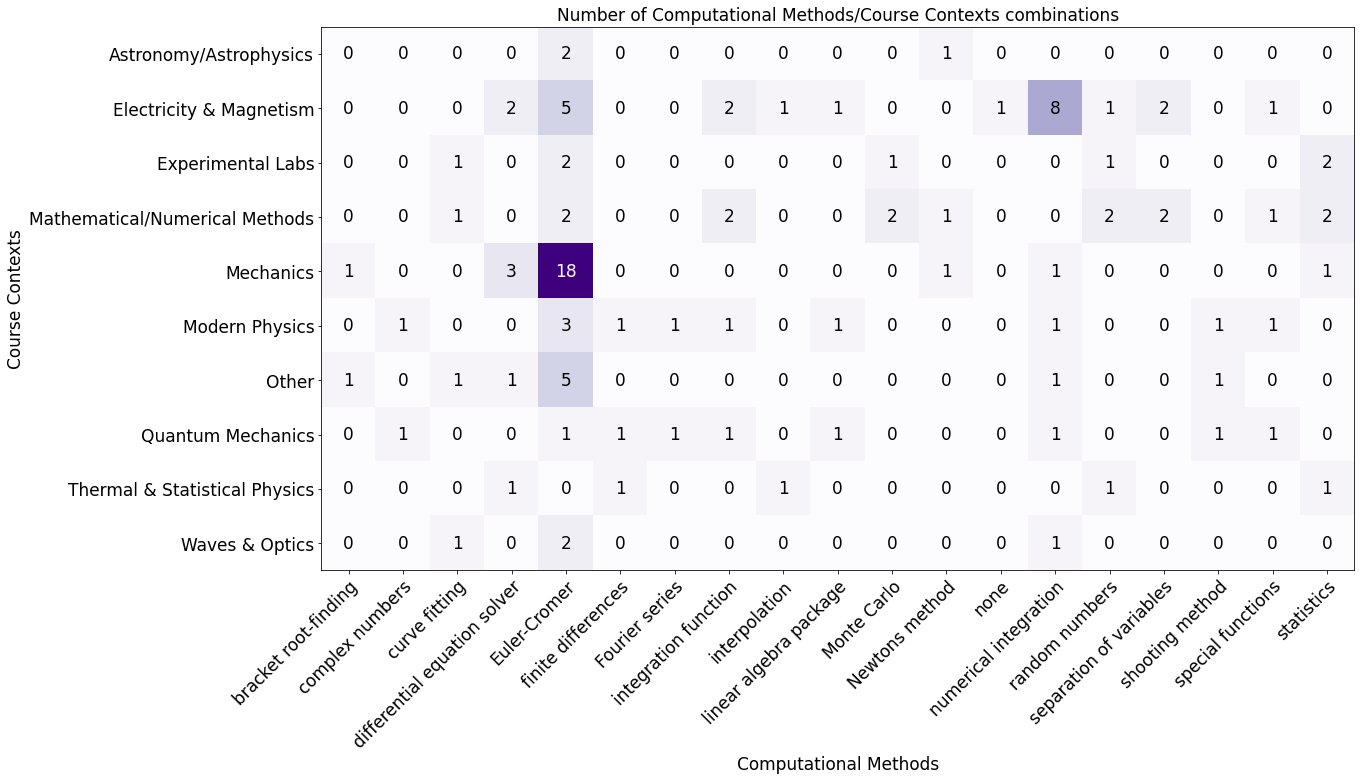}
\caption{\label{fig:SubjectMethodheatmap} Frequency with which each combination of Course Context and computational method occurs in the PICUP Collection.}
\end{figure*}

Different Course Contexts in physics lend themselves more readily to different computational methods, depending on the underlying mathematical principles or historical treatments of the subject. Figure \ref{fig:SubjectMethodheatmap} shows the number of times each combination of Course Context and computational method occurs in the PICUP Collection. By far, the most frequent combination (in 18 ESs) is found between Mechanics and the Euler-Cromer method. This high frequency is unsurprising, as much effort has been devoted in the PICUP community to the first-semester introductory mechanics course, and the Euler-Cromer method is a simple means of solving the differential equations that govern classical mechanics. The second most frequent combination (in 8 ESs) is found between Electricity \& Magnetism and numerical integration. Again, this high frequency is unsurprising, given the number of ESs designed for use in second-semester introductory Electricity \& Magnetism.

The Euler-Cromer method also proves popular in the Electricity \& Magnetism context (5 ESs) and in 5 ESs classified as Other under Course Context. The remaining combinations of Course Context and computational method occur infrequently (1-3 ESs) or not at all. These ``missing'' combinations offer potential for further development, which we discuss in the next section.

% \subsection{Course Contexts and Platforms}

% An examination of the combinations of Course Context and platform gives insight into which platforms ES authors find most compatible with the types of activities conducted in each Course Context. Figure \ref{fig:SubjectPlatformheatmap} indicates that, in the popular Mechanics Course Context, many platforms are frequently used, from long-standing environments spreadsheets and Fortran to in-browser editors such as GlowScript and Jupyter. In Electricity \& Magnetism, there is a more centralized preference for IPython. This shift perhaps indicates a widespread need to offer plenty of entry options to first-semester physics students who may be programming for the first time, with this need being less pronounced in the second semester. The remaining combinations of Course Context and platform seem to indicate no preference beyond the limited number of example codes provided in many ESs as discussed in Section \ref{subsec:platforms}.

% \begin{figure}[tb]
% \includegraphics[width=0.45\textwidth]{Subject-Platform_heat_map.png}
% \caption{\label{fig:SubjectPlatformheatmap} Frequency with which each combination of Course Context and platform occurs in the PICUP Collection.}
% \end{figure}

\subsection{Course Contexts and Engagement Elements}

\begin{figure*}[tb]
\includegraphics[width=\textwidth]{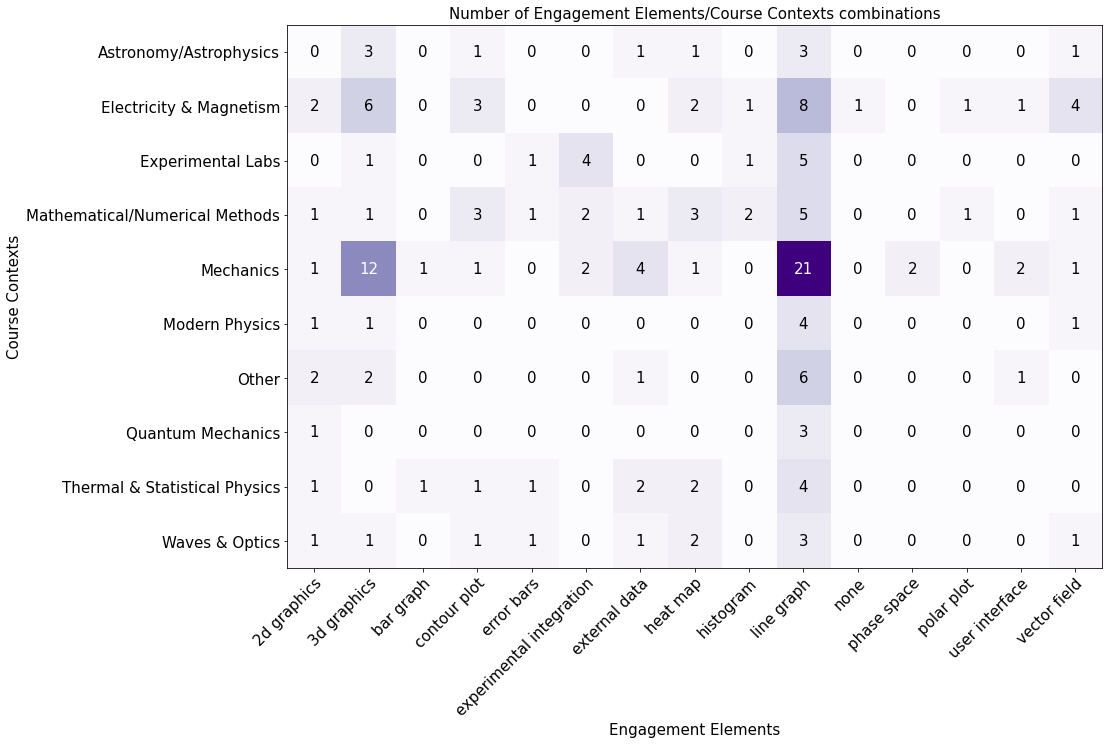}
\caption{\label{fig:SubjectEngagementheatmap} Frequency with which each combination of Course Context and engagement element occurs in the PICUP Collection.}
\end{figure*}

The engagement elements discussed in Subsection \ref{subsec:ee} are all useful in promoting student participation and learning. It is worthwhile to examine how frequently these engagement elements appear across the various Course Contexts to see where they could be used more frequently. Figure \ref{fig:SubjectEngagementheatmap} shows the number of times an engagement element is used in an ES from each Course Context. Line graphs appear in every Course Context, as they are the most common representation of data in a physics class. We also see prevalent use of three-dimensional graphics in Mechanics (largely owing to the use of GlowScript for easy three-dimensional graphics) and in Electricity \& Magnetism (due to the use of GlowScript for three-dimensional animations and Matplotlib for three-dimensional graphs). Experimental integration is used most frequently in the Experimental Labs Course Context, but is rarely employed elsewhere, even though most of the Course Contexts lend themselves to undergraduate-level laboratory work.

% \subsection{Computational Methods and Platforms}

% not much to learn here... 

\subsection{Computational Methods and Engagement Elements}

\begin{figure*}[tb]
\includegraphics[width=\textwidth]{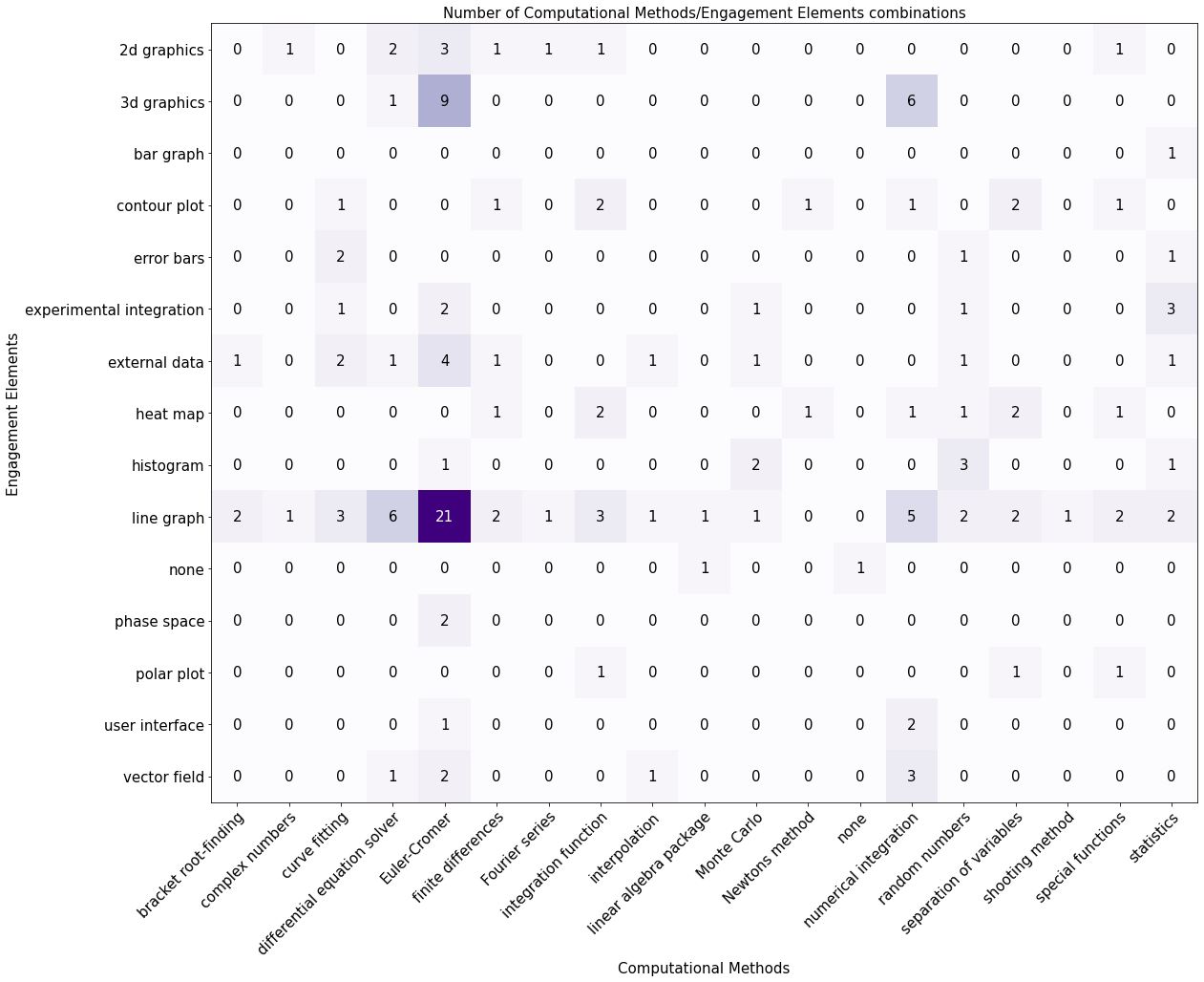}
\caption{\label{fig:MethodEngagementheatmap} Frequency with which each combination of computational method and engagement element occurs in the PICUP Collection.}
\end{figure*}

Many computational methods tend to be associated with a specific type of output, which lends itself to particular visualizations. For example, a histogram is a natural choice when viewing outcomes from a process involving randomness, but one rarely views data from a deterministic process in a histogram. We can see such relationships in Figure \ref{fig:MethodEngagementheatmap}, which shows the number of ESs that use each possible combination of computational method and engagement element. The most frequent combination occurs between Euler-Cromer (the most popular computational method) and line graphs (the most popular engagement element), followed by Euler-Cromer and 3d graphics (largely owing to the use of GlowScript for easy three-dimensional graphics). Many engagement elements are used with only a few computational methods, indicating room for further development in future ESs.

\subsection{Platforms and Engagement Elements}

\begin{figure*}[tb]
\includegraphics[width=\textwidth]{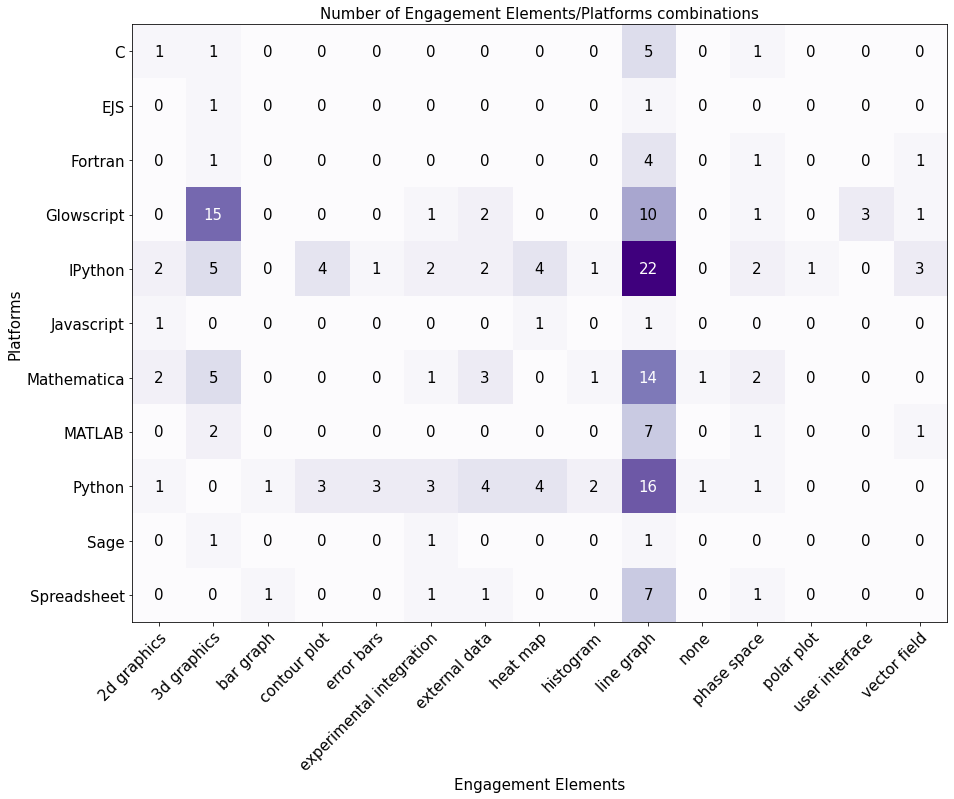}
\caption{\label{fig:PlatformEngagementheatmap} Frequency with which each combination of platform and engagement element occurs in the PICUP Collection.}
\end{figure*}

Some programming platforms make it easier to incorporate engagement elements, as seen in Figure \ref{fig:PlatformEngagementheatmap}. We see that most kinds of graphs and plots can be found in ESs that use IPython, due to the wealth of visualization resources available through Matplotlib and the report-like environment created in an IPython notebook. Line graphs are used across every platform. Three-dimensional grpahics are used most frequently with GlowScript, with some use in IPython and Mathematica. 

\section{Evaluation and Recommendations}

Now that we have examined the contents of the PICUP Collection, we evaluate the Collection's strengths and areas in which it could be further developed. We conclude this section with recommendations for ES authors to consider.

\subsection{Strengths of the PICUP Collection} % Add figure feferences.

The PICUP Collection has developed a strong representation of Course Contexts and topics appropriate to the First Year and Beyond the First Year Course Level. More than one-third of ESs are listed under both of these Course Levels, indicating a versatility of ESs that are approachable at the introductory level but still retain learning value at the sophomore level and beyond. The majority of ESs can be completed within a typical class or lab period.

The ESs of the PICUP Collection also tend to be focused in topic and scope, with most ESs addressing 3 or fewer conceptual topics and implementing only 1 computational method. This focus can help instructors choose an appropriate ES for their course context without incurring significant additional cognitive overhead for their students \cite{Redish:93}. 

Many PICUP ESs endeavor to realize the promise that computational physics activities grant students access to analytically intractable problems \cite{Redish:93}. This effort is demonstrated in the high frequency of activities based on gravitation, which is touched on in introductory mechanics courses but not fully explored due to its analytical challenges. Such activities offer greater insight into the topic and demonstrate to students the power of computation as an avenue of physics research.

The ESs make repeated use of a handful of computational methods approachable at the first-year level. These frequently used methods are selected based on their mathematical fit with first-year mechanics and electricity \& magnetism. Having a common set of frequently used methods across multiple ESs helps first-time adopters by lowering the learning demand they experience during preparation and delivery. Additionally, an instructor adopting multiple ESs that use the same computational method reinforces student learning throughout the semester and enables students to increase their focus on higher-level computational thinking and physics learning throughout the semester.

The ESs tend to make frequent use of a common set of programming platforms that are known to be accessible to novices. This accessibility is important to maintain student motivation and engagement, especially as they learn physics and programming in tandem. Additionally, nearly all ESs are platform-agnostic, allowing instructors to easily adapt them based on their own familiarity or the programming context at their institution.

We also find that the ESs have demonstrated the use of various visualization and interactive elements likely to help engage students in the processes of analysis, reflection, and synthesis. Some computational methods lend themselves more readily to different engagement elements, and some programming platforms enable a greater variety of engagement elements. These relationships are visible in the ESs' use of engagement elements.

\subsection{Areas for Development} % Add figure feferences.

With these strengths established, PICUP is poised to expand the Collection in a few key areas.

Perhaps most notably, there is a lack of ESs at the High School and Advanced Course Levels. With  programming and computational thinking being prioritized in high school \cite{NGSS}, and graduate students expected to conduct computational research activities \cite{GradResearch} that may outpace their undergraduate training, there is a strong need for computational resources at these Course Levels. Existing ESs could be adapted to serve these populations, such as by easing the technical rigor of First Year ESs to make them accessible to High School students, and by extending the analysis required in Beyond the First Year ESs to make them appropriate for the Advanced level. It is also worthwhile to note that, while the qualities of a High School and First Year context are generally clear, the distinction between Beyond the First Year and Advanced is not clearly articulated, and likely depends on institutional context. There is probably much room for development of ESs cross-listed between Beyond the First Year and Advanced.

We also see that there are many Course Contexts (Condensed Matter Physics, High Energy/Particle Physics, Biophysics, and Chemical/Molecular Physics) currently unrepresented in the PICUP Collection. (While finalizing this article, the author did successfully publish an ES for Condensed Matter Physics, although it is not included in the analysis here.) Given that these subjects actively involve ongoing computation-based research, it would be beneficial for educators to develop ESs to help students begin learning these research practices. Challenges associated with incorporating one's own research methods into coursework is a topic of conversation in the PICUP community, and warrants further investigation.

Another significant need is found in the low number of programming platforms employed in each ES (Figure \ref{fig:Platformbargraph}). Despite the editors' recommendation that each ES include code templates and completed codes for at least three different platforms, over two-thirds of the ESs use only one platform. Additionally, many that do include multiple platforms use different implementations of the same language (such as Python and IPython). The ESs would be more broadly useful if additional implementations were submitted. Since nearly all ESs are platform-agnostic (Figure \ref{fig:AgnosticCountbargraph}), this expansion should be a straightforward process.

While significant work has been accomplished in a few favored topics, there are many topics underexplored in the PICUP Collection. For example, ESs involving gravitation could easily be extended to explore interatomic interactions using the Lennard-Jones force, and yet this application remains absent in the PICUP Collection. Similarly, while there are many ESs about electrostatics, there are very few about magnetostatics, even though they could follow a complementary structure. 

Physics academic programs tend to approach their subjects in isolation. Course Contexts such as quantum mechanics and electricity \& magnetism are learned in their separate courses and are allowed to interact very infrequently. This isolation is highlighted by the infrequent combination of Course Contexts among PICUP ESs (Figure \ref{fig:SubjectHeatMap}), with Subject combinations primarily occurring in Advanced-level ESs. While physics subject isolation is traditionally necessitated by the technical challenges associated with combining the subjects' disparate formalisms, computational activities offer the promise of reducing such technical challenges. More ESs could be developed to help students explore connections between physics subjects. 

While the current PICUP ESs make excellent use of a few computational methods such as Euler-Cromer and numerical integration, these are certainly not fully representative of the type of methods used in physics research and the broader world of STEM. For example, curve fitting and statistics are underutilized in the PICUP collection, despite being immensely important in the world of data science and analytics. Likewise, no one can deny the importance of linear algebra in physics and engineering research, and yet only a few PICUP ESs make use of linear algebra libraries or algorithms. Many other methods (such as root-finding, random numbers, and constructing series) that are frequently used in upper-level physics curricula and research could be better represented, as well.

Future ESs could also explore the application of computational methods to new Course Contexts. For example, curve fitting and interpolation are currently unused in over half the Collection's Course Contexts, even though every field of physics uses these practices. Numerical integration, despite its strong presence in Electricity \& Magnetism, is used infrequently across other Course Contexts.

The fact that most ESs use a single computational method helps to maintain a low cognitive overhead for instructors and students, but it would be helpful to see ESs that integrate multiple methods for comprehensive assignments or Advanced-level activities. Indeed, Langbeheim et al recommend giving students opportunity to try applying multiple models to a single problem \cite{langbeheim2020webbing}.

The current ESs provide excellent examples of the use of various engagement elements. Future authors would do well to incorporate underutilized engagement elements, especially visualization formats that are growing in demand such as heatmaps and three-dimensional graphics \cite{visualization}. Additionally, using multiple engagement elements in each ES would help capture the attention of more students and reinforce learning across a greater range of Course Contexts.

\subsection{Recommendations for Authors}

In summary, we present the following recommendations for authors submitting to the PICUP Collection:

\begin{enumerate}

\item \textit{Replicate the successes of the current ESs at the High School and Advanced Course Levels.} With the existing ESs as a guide, authors can extend best practices in ES development to these underrepresented contexts. A greater number of High School-level ESs would help prepare more students for undergraduate-level STEM coursework, and Advanced-level ESs would extend PICUP's successes throughout the undergraduate level and into the graduate level. Since the High School designation was only recently added to the list of Course Levels in the Collection, it is possible that some existing First Year ESs are also appropriate for a High School context (or could easily be adapted for High School). Focus could also be devoted to encouraging and equipping high school teachers to develop ESs appropriate for their context.

\item \textit{Expand the representation of programming platforms.} A greater variety of platforms would support more instructors with varying personal experience and institutional contexts, and would expose students to the various programming languages they are likely to encounter in the professional world. PICUP accepts additional code templates and completed codes from users after an ES is published, so physics educators with a background in underutilized programming languages can provide additional support for existing ESs. As most ESs are written by a single author, future ESs would benefit from collaborative efforts to provide codes for multiple platforms at the time of submission.

\item \textit{Develop ESs around more context-rich, analytically intractable problems.} The success of gravitation as a topic among the ESs indicates the power of computational physics activities to explore a rich parameter space in the context of an analytically intractable problem with direct real-world applications. Efforts to develop ESs about problems with similar features (such as molecular dynamics, non-linear circuit elements, electrodynamics, and time-dependent quantum mechanics) would enrich the computation-based learning opportunities available to students. These topics could extend into Course Contexts that currently have no ESs in the Collection.

\item \textit{Develop ESs that integrate Course Contexts that are typically isolated from each other.} With existing ESs as building blocks, authors can explore topics at the intersection of Course Contexts, such as studying magnetic forces with Lagrangian mechanics or comparing classical and quantum models of conduction. Additionally, it would be beneficial to develop ESs that extend physics concepts and computational methods to applications in other fields frequently of interest to students, such as biology, chemistry, or engineering.

\item \textit{Include a greater variety of computational methods and engagement elements, especially those that students are likely to encounter in graduate education and industry.} Underutilized methods include regressions and curve fitting, linear algebra, random number generation, and data science. Underutilized engagement elements include heat maps, histograms, phase space plots, and polar plots.

\end{enumerate}

\section{Conclusion}

We have presented an analysis of the contents of the PICUP Collection of Exercise Sets. This Collection is a valuable resource in the endeavor to incorporate computational activities into physics education. We have observed that the PICUP Collection tends to focus on undergraduate-level needs across a variety of subjects (most frequently mechanics and electricity \& magnetism) using a variety of computational methods (most frequently the Euler-Cromer method and numerical integration). The Collection provides templates and example codes in a variety of programming languages and environments (with Python being the most frequently implemented) and using these languages to introduce a variety of engagement elements. To strengthen the Collection, we outline several areas for development and make specific recommendations for authors of future Exercise Sets. 

We thank the Exercise Set authors and PICUP's leaders, editors, and reviewers for their invaluable contributions to physics education.

%%% BRIAN STOPPED HERE %%%

\nocite{*}

\bibliography{apssamp}% Produces the bibliography via BibTeX.

\end{document}